# Data Findability, Governance, and Community Engagement for Māori Research Data Sovereignty.

*Paul T. Brown[1], Kiri West[2], Maree Sheehan[3], Hana Rapata[2], Te Taka Keegan[1].*

[1]University of Waikato, Hamilton, NZ.
[2]University of Auckland, Auckland, NZ.
[3]Te Wānanga o Aotearoa, Hamilton, NZ.

## Abstract

Māori data sovereignty (MDSov) has established important principles for recognising Māori rights and interests in data. Relatively little attention has been given to how these principles can be operationalised within research institutions who, as a function of their operations, collect and use Māori data. This paper introduces the concept of Māori Research Data Sovereignty (MRDSov), extending existing understandings of MDSov into the specific context of research data and the research data lifecycle. Drawing on Indigenous Data Sovereignty scholarship and research data management literature, we define Māori research data and position MRDSov as the application of MDSov principles to research data produced by, about, or for Māori. We argue that operationalising MRDSov requires three interdependent elements: data findability, data governance, and community engagement. Data findability enables Māori research data to be identified and contextualised; governance provides mechanisms through which Māori authority over research data can be exercised; and community engagement grounds both in enduring relationships with Māori communities. We further demonstrate how these elements collectively enact the principles of MDSov within institutional research settings, providing a practical pathway for aligning research data practices with Te Tiriti o Waitangi obligations and Indigenous data governance expectations. The paper contributes a conceptual framework for universities and other research organisations seeking to embed Māori authority, accountability, and relationships throughout the research data lifecycle.

### Keywords

Indigenous data sovereignty, Māori research data sovereignty, data findability, data governance, community engagement.

## 1. Introduction

Despite being framed as a 'net positive' for society, research is not neutral and often does little to ensure tangible positive outcomes for Indigenous Peoples. Western research methods and methodologies are often extractive (Tuhiwai Smith, 2021; Godrie, 2025), cause harms (Garrison, 2013; Mello and Wolfe, 2010; Rewi *et al.*, 2022), and deficit framed (Andersen and Walter, 2013). In an era where the ‘datafication’ of the world (Meijas and Couldry, 2019) is driving new technologies that make research more efficient, these systems continue to structurally exclude Indigenous peoples.

Western research institutes such as Universities and crown research institutes are a large contributor to the Aotearoa New Zealand (hereafter referred to as NZ) research landscape, and therefore collect and store significant amounts of data (Sterling *et al.,* 2023). These institutes have had a long history of collecting data about Māori, the Indigenous peoples of NZ. However, these processes are generally conducted without significant consideration for how this data should be handled, managed, used, and how the distribution of benefits from the knowledge generated is given back to the communities. Policies upholding Indigenous rights to the research data about Māori that Universities collect are scarce. To aid in this ideological and policy gap, we present the idea of Māori research data sovereignty (MRDSov), an exploration that expands on the ideas of Māori data sovereignty (MDSov) in the specific context of Māori research data.

The paper first reviews the literature on Indigenous Data Sovereignty (IDSov), Māori Data Sovereignty (MDSov), and the research data landscape in Aotearoa New Zealand. It then defines Māori research data and introduces Māori Research Data Sovereignty (MRDSov) as the application of MDSov to the research data lifecycle. Building on these foundations, the paper proposes three interdependent elements for operationalising MRDSov: data findability, data governance, and community engagement. It then examines how these elements work together to operationalise MRDSov and collectively give effect to the principles of MDSov. The paper concludes by considering the implications of this framework and recommending a pathway for universities and other research institutions seeking to implement MRDSov.

The purpose of this paper is to establish a conceptual foundation for Māori Research Data Sovereignty within the context of research institutions. To our knowledge, this is the first paper to conceptualise Māori research data as a distinct domain of Māori data, define MRDSov as the application of MDSov to the research data lifecycle, and propose an operational framework that reframes data findability, data governance, and community engagement as interconnected mechanisms for enacting MDSov in research practice. In doing so, the paper introduces a conceptual language through which Māori research data can be understood and operationalised within research institutions.

# 2. Māori Research Data Sovereignty

## IDSov, MDSov, and Universities

IDSov challenges conventional understandings of data sovereignty (DSov) by asserting that Indigenous data should be subject to the laws of the Indigenous peoples to whom the data relates (Carroll *et al*., 2020; Rainie *et al*., 2019; Walter & Suina, 2019), as opposed to the nation-state where it sits. IDSov provides a basis for challenging longstanding power imbalances by giving Indigenous peoples practical ways to assert authority over data generated and used by governments and corporations (Cormack and Kukutai, 2022). Indigenous nations and collectives have developed their own distinct expressions of IDSov, including the Maiam nayri Wingara principles (Maiam Nayri Wingara, 2018), Pasifika data sovereignty in NZ (Pacific Data Sovereignty Network, 2021), the First Nations OCAP® Principles (FNIGC, 2024), emerging approaches to data sovereignty across the Sámi nations (Carrasco et al., 2025), and the Métis Nation Data Strategy (Métis National Council, 2025). A growing body of work also continues to demonstrate how IDSov can be operationalised. For example, see Carrol *et al*., (2021); Kukutai and Taylor, (2016); Walter *et al*., (2021); Wilks (2018).

Māori data are considered a *taonga*, a treasured possession that requires a level of mana (respect) and active protection (Hudson *et al.*, 2017; Ruckstuhl, 2023; Waitangi Tribunal, 2011). Māori data is generally defined as data about Māori peoples, Māori culture, or environments that Māori have rights or interests in. MDSov argues that Māori data should be subject to the laws and governance structures of Māori. Several frameworks exist for the appropriate governance and use of Māori data, including the MDSov principles which present high-level Māori concepts for which MDSov is built upon (Te Mana Raraunga, 2018). This document outlines six principles and 16 sub-principles which advocate for Māori to have sovereignty over their data, and provides a basis for developing practical pathways to operationalise MDSov. We provide the reader a brief overview of the high-level MDSov principles in Appendix A. Also, see Brown *et al*., (2024) for a discussion about how these concepts also relate to Te Tiriti o Waitangi/The Treaty of Waitangi, and how they relate to the principles of the treaty. For more on the corresponding sub-principles for each principle, see Te Mana Raraunga (2016).

The Māori data governance model (Kukutai *et al*., 2023a) builds on the MDSov principles to provide further guidance over the governance of Māori data. The Māori Data Sovereignty and Privacy framework (Kukutai *et al*., 2023b) extends the principles of MDSov into the field of privacy. In addition, the Māori algorithmic sovereignty framework (Brown *et al*., 2024, O'Neale, 2025), reframes the MDSov principles to cover the context of algorithmic systems, and provides a framework for the critical analysis of operational algorithms from a Māori perspective. Ngā Tikanga Paihere (Stats NZ, 2020) is used by Statistics NZ/Tatauranga Aotearoa, NZ's official data collection agency, to guide the cultrually appropriate use of Māori data in the Integrated Data Infrastructure (Milne, *et al*., 2019). Māori tribal groups known as Iwi and Hapū (subgroups of Iwi) have brought the ideas of MDSov "to-the-ground" and express and practice their own ideas of data sovereignty. See for example Gifford, (2019); Hudson, (2016); Hudson *et al*., (2016); Kukutai, (2024).

A paper regarding Indigenous research sovereignty and the relationship to research data, by Hudson *et al*., (2023), argued that Indigenous sovereignty should also extend to data and research ecosystems by recognising and operationalising IDSov. The authors provide a suite of specific rights in relation to data, such as Indigenous rights to self-determination, consent, governance, privacy, and benefits. These rights are essential to ensure that research data practices align with Indigenous aspirations and prevent the misappropriation or misuse of Indigenous knowledge and data. Marley (2020) outlined the role of universities in upholding IDSov over data relating to American Indians, Alaskan Natives and Indigenous Peoples of North America, and provided some of the preliminary steps that the Universities of Oregon, New Mexico, and Montana have taken. A communique from the Global Indigenous Data Alliance (Prehn, 2023) urged universities to take concrete steps to uphold IDSov and Indigenous Data Governance (IDGov; Lovett *et al*., 2019). Institutions are called to recognise that all data concerning Indigenous peoples is Indigenous data, identify and account for such data held within their systems or by research partners, and implement policies that ensure Indigenous authority, access, and control over its present and future use. Universities are encouraged to support Indigenous leadership in assessing and managing digital research infrastructure, embed IDSov and IDGov principles throughout data management strategies and operational plans, provide training for administration, faculty, and students, and allocate sufficient resources to enable Indigenous communities to govern their data on their own terms.

## Research Data

Research data refers to data that are systematically collected, generated, or analysed for the explicit purpose of answering research questions, testing hypotheses, or producing new knowledge, and are created within defined methodological, ethical, and analytical frameworks (OECD, 2015). This distinguishes research data from general data, which are often produced through administrative or operational activities and are not necessarily designed to be reusable beyond an operational context. Research data derive their value from the standards and practices that enable others to understand how the data were produced and what claims they can support (Borgman, 2015). This makes research data important for decision-making, as they provide an evidence base that allow hypotheses to be examined. Leonelli (2016) states that (scientific) data become reliable for decision-making only when they are appropriately curated and situated within their epistemic context, highlighting the role of research data as a foundation for knowledge production.

The authors of Wilkinson *et al.,* (2016) presented the FAIR principles (Findable, Accessible, Interoperable, and Reusable) as a set of guiding principles to ensure that research data retain value beyond their initial use. They argue that research data lose their scientific and societal value because they cannot be reliably found, accessed, or linked (with other data and datasets) once projects end, particularly in data-intensive and computational research environments. The authors emphasise that data become valuable not only by being stored or shared, but by being accompanied by unique identifiers, informative metadata, and standardised formats that allow both humans and machines to understand and reuse them. Perhaps most importantly, they stress that FAIR data does not require the data to be open, but rather that conditions to its access are clearly defined. This enables reuse under appropriate governance arrangements. FAIR principles are essential because they transform research data from isolated project outputs into assets that support reproducibility, knowledge building, and long-term decision-making across disciplines.

The CARE Principles for IDGov provide a complementary framework to FAIR by emphasising the Collective Benefit, Authority to Control, Responsibility, and Ethics of data related to Indigenous peoples (Carroll *et al*., 2020). While FAIR focuses on the technical aspects of the reuse of research data, CARE balances this with the ethics of data use, ensuring that data practices respect Indigenous values. CARE recognises that data about Indigenous communities is not neutral and may have implications for Indigenous self-determination. By integrating CARE with FAIR, researchers and institutions can ensure that data are not only technically reusable but also governed in ways that are ethically responsible and culturally appropriate, supporting both robust scientific use and the rights and priorities of Indigenous communities (Carroll *et al*., 2020; Rainie *et al*., 2019). In practice, CARE provides guidance on ensuring that FAIR data infrastructures, metadata standards, and sharing policies incorporate authority, benefit, and ethical stewardship, enabling research that enhances IDSov.

A piece on research data in NZ by Sterling *et al.*, (2023) describes research data as a critical strategic asset for the research and innovation ecosystem of NZ. However, when analysing the landscape of NZ's research data, the authors found that much of the data generated or used by researchers is not effectively managed according to expectations. They note that significant portions of research data are not FAIR, which limits their value for future research and decision-making, and that there are systemic gaps in how research institutions steward data across the lifecycle. In relation to Māori data, the authors highlight that data "about, from, or connected to Māori" frequently fails to meet the requirements of MDSov frameworks or the CARE Principles for IDGov, indicating shortcomings in recognising Māori governance preferences in research data practice. This lack of alignment has important implications for universities and other research organisations. To address these issues, institutions must better integrate constitutional and Te Tiriti o Waitangi obligations, Māori data governance, and culturally grounded practices into research data policies, infrastructure, and organisational culture to support equitable research outcomes and meet national expectations. The report, in 20 recommendations, urges universities and the broader research sector to adopt systemic improvements including embedding MDSov considerations, strengthening metadata standards, and upgrading infrastructure to enhance research data quality and value for all communities.

## Conceptualising Māori Research Data Sovereignty

Building on the definition of research data, Māori research data can be defined as research data that are about Māori people, communities, knowledge systems, resources, or environments, or that are generated by or for Māori through research activities, and that carry geneological links and cultural significance.

MRDSov refers to the inherent right of Māori to exercise authority and responsibility over Māori research data across the entire research data lifecycle, in ways that uphold the principles of MDSov. Building on the definition of Māori research data, MRDSov recognises such data as a taonga, not just an output of research. It affirms that decisions about the collection, governance, access, analysis, interpretation, sharing, and reuse of Māori research data must be determined in conjunction with Māori, according to Māori values. Under this framing, MRDSov ensures that appropriate research methods are matched with appropriate Māori data governance and protection, so that research data supports knowledge creation and contributes to collective benefit and the protection of Māori rights and interests.

In this work, Māori research data and MRDSov are defined by extending existing understandings of Māori data and MDSov into the specific context of research, rather than by redefining them or introducing new principles. Māori research data could be understood as a subset of Māori data. MRDSov therefore represents the application of MDSov to the research data lifecycle. Because MRDSov is grounded in MDSov, the underlying principles do not change, they are just operationalised within research contexts. The core MDSov principles along with their associated sub-principles remain fully relevant, and should guide decision-making and practice. What changes is not the foundation, but the domain of application.

# 3. Operationalising MRDSov – Three Key Elements

Operationalising MRDSov requires intentional and critical research practices that engage research and data management practice within the theories and frameworks described above. For Māori research data, sovereignty must be enacted across the full data lifecycle, from creation to reuse and ultimately return to the communities from which it derives, or destruction. In practice, this can be achieved through three interrelated elements: data findability, data governance, and community engagement. Contextualising these elements within the framework of MDSov, helps to ensure that Māori research data are visible and usable on Māori terms, governed according to Māori values and Māori authority, and grounded in enduring relationships with Māori communities. This approach aligns MDSov principles with IDGov frameworks such as the MDGov model and CARE, while also engaging constructively with modern research data infrastructures. Operationalising these elements requires research infrastructure capable of identifying, managing, governing, and supporting Māori research data throughout the research data lifecycle.

## Element 1 - Findability

Data findability refers to the ability to locate, identify, and understand the existence and nature of research data through appropriate descriptions, metadata, and discovery mechanisms. In the context of MRDSov, it may be best understood as both a mechanism for upholding mana and rangatiratanga over Māori research data, and also as a technical function of discovery systems. Māori research data are genealogically connected to people, their places and their histories. It cannot be properly collected or meaningfully governed, interpreted, and used if their existence, origins, or conditions of use are not acknowledged.

Universities in NZ lack the internal mechanisms to track Māori research data within their data ecosystems, and therefore have no idea what data they have, or how much actually exists. This lack of findability creates an information vacuum, where neither the institution, nor Māori can account for, and govern, Māori research data. Borgman (2015) argues that data derive their value from the contextual information that allows others to understand how and why they were produced. For Māori data, this context includes tikanga (Māori culture, values, and protocols), provenance, and collective authority. The OECD (2015) identifies findability and metadata as prerequisites for research data to function as enduring scholarly and policy assets, yet MDSov requires that such metadata also signal Māori interests and responsibilities, not just technical descriptors.

Carroll *et al.* (2021) demonstrate how Indigenous data governance can be operationalised within data systems through culturally informed metadata. Liggins *et al*., (2021) provides a practcal tool to include labels and notices that communicate Indigenous authority and expectations of use. In this way, data findability can support active protection by ensuring Māori research data remain visible, and enables appropriate governance, informed decision-making, and the protection of Māori rights and interests across the data lifecycle.

## Element 2 – Data Governance

Data governance is the primary mechanism through which MRDSov is enacted, as it determines who has authority to make decisions about the research data. The MDGov model positions governance not simply as policy compliance, but as a system of collective authority, accountability, and stewardship grounded in rangatiratanga, whakapapa, and kaitiakitanga (Kukutai *et al.*, 2023a). Within this framework, Māori research data are recognised as assets whose governance must reflect Māori values, aspirations, and obligations to both current and future generations.

Governance therefore requires clear decision-making roles, recognition of Māori authority at appropriate scales (for example, iwi, hapū, or Māori collectives), and mechanisms that ensure data use aligns with agreed purposes and benefits, and given with prior and informed consent. This approach contrasts with generic data governance models, which often prioritise legal ownership or administrative control. Embedding this form of Māori governance within research data systems is key to operationalising MRDSov, as it ensures that research data practices uphold Māori rights and interests while enabling ethical and accountable use of data for research and decision-making. In NZ, the role of universities should focus on building governance structures that facilitate MDSov over research data, as the sovereignty of Māori data is an exclusive right of Māori that institutions can support but never themselves claim to possess.

## Element 3 – Community Engagement

Community engagement refers to reciprocal relationships between researchers, institutions, and Māori communities throughout the research data lifecycle. Within MRDSov, engagement is not a one-off consultation but an ongoing process of partnership that shapes research questions, data practices, interpretation, and outcomes. Meaningful engagement ensures that Māori research data are generated and used in ways that reflect community priorities, expectations, and definitions of benefit, rather than extractive research models that separate data and research from people. Both MDSov scholarship and Indigenous research ethics literature emphasise that strong relationships are foundational to ethical data stewardship and collective benefit (Kukutai & Taylor, 2016; Tuhiwai Smith, 2021). Sterling *et al.*, (2023) reinforces this by highlighting the need for culture change within universities, where engagement with Māori must be embedded in research data practice if data are to be trusted, legitimate, and socially valuable. In this way, community engagement operationalises MRDSov by grounding technical and governance mechanisms in relationships and shared responsibility.

## Relationships Between Findability, Governance, and Engagement

Data findability and data governance are mutually reinforcing within MRDSov. This is because governance cannot be meaningfully exercised over data that are not findable. Findability provides the informational infrastructure that allows Māori research data to be identified, linked to their provenance, and associated with culturally grounded conditions of use. In the absence of findability, governance defaults to generic institutional norms that risk eroding rangatiratanga. Data are only governable when their context is preserved; for Māori research data, this context includes whakapapa and tikanga. Findability therefore enables governance by supporting informed, accountable decisions about access, reuse, and kaitiakitanga across the research data lifecycle.

Findability is also closely tied to community engagement, as Māori communities cannot engage with, benefit from, or exercise authority over research data that they cannot locate or recognise. Within MRDSov, findability is not oriented toward unrestricted openness, but toward transparency on terms defined by the Māori individuals and collectives from which the data comes from. Indigenous data scholarship demonstrates that culturally informed metadata and findability mechanisms can make data visible while signalling ethical obligations and governance expectations (Carroll *et al*., 2021, Liggins *et al*., 2021). When Māori research data are findable in ways that reflect relationships and whakapapa, communities will be better positioned to participate in interpretation, challenge misrepresentation, and contribute to decisions about future use, reinforcing whanaungatanga and kotahitanga.

Data governance and community engagement are likewise inseparable, as Māori data governance frameworks emphasise collective authority, accountability, and relational decision-making rather than individual consent alone. The MDGov framework conceptualises governance as an ongoing practice grounded in engagement with Māori collectives and responsive to their aspirations (Kukutai *et al.,* 2023a). Governance structures that recognise Māori authority create meaningful pathways for engagement, while sustained engagement ensures that governance arrangements remain grounded in tikanga. Together, governance and engagement operationalise rangatiratanga and manaakitanga within institutional research settings.

Taken together, data findability, governance, and community engagement operate as interdependent pillars that collectively operationalise MRDSov. Emerging empirical evidence and a growing body of conceptual scholarship, suggests that IDSov can strengthen ethical research practice by enhancing trust between researchers and Indigenous communities, and support the translation of Indigenous sovereignty principles into institutional data governance and policy. In this way, MRDSov can be implemented through the alignment of infrastructure, governance, and relationships across the research data lifecycle.

## Relationships With MDSov Principles

The operationalisation of Māori Research Data Sovereignty (MRDSov) through data findability, data governance, and community engagement can be understood as the practical enactment of MDSov principles within research systems. Each operational component translates foundational Māori values into institutional and infrastructural practices that shape how research data are created, managed, and used, see Appendix B.

Data findability most directly gives effect to the principle of whakapapa, as it requires that Māori research data be visibly connected to their origins, relationships, and contexts. Through culturally informed metadata and provenance records, findability ensures that Māori research data are not rendered abstract or decontextualised but remain embedded within the relational networks that give them meaning. In doing so, findability also supports rangatiratanga by signalling Māori authority, governance expectations, and conditions of use at the point of discovery, enabling Māori decision-making to be recognised before data are accessed or reused. This visibility is a precondition for kaitiakitanga, as long-term stewardship responsibilities cannot be exercised over data that are invisible, poorly described, or disconnected from their whakapapa.

Data governance, as an operational principle, most explicitly enacts rangatiratanga by establishing who has the authority to make decisions about Māori research data and on what basis. Governance frameworks grounded in Māori values shift decision-making away from default institutional ownership models toward collective authority. Within this framing, kotahitanga is expressed through governance structures that align data practices with shared Māori aspirations and collective benefit, while manaakitanga is operationalised through processes that ensure Māori research data are used ethically, respectfully, and in ways that protect Māori wellbeing, reputation, and rights. Kaitiakitanga is further reflected in governance arrangements that emphasise intergenerational responsibility across the research data lifecycle.

Community engagement operationalises whanaungatanga by centering relationships between researchers, institutions, and Māori communities as the foundation of legitimate research data practices. Engagement ensures that Māori are active participants in shaping how research data are collected, interpreted, governed, and used. It also ensures that Māori are not just the subject of the data and research. Through consistent and sustained engagement, Māori sovereignty can be strengthened, by aligning research objectives and data uses with collective Māori goals and by creating meaningful pathways for Māori communities to influence decisions about their data. Manaakitanga is expressed through respectful, reciprocal engagement practices that recognise Māori knowledge, time, and contributions, and that mitigate harm arising from misrepresentation or extractive research practices. Engagement also strengthens kaitiakitanga by enabling communities to exercise ongoing oversight and care over how data relating to them are stewarded and reused over time.

As stated previously, these operational components do not function independently; rather, they are interdependent and mutually reinforcing. Findability enables governance by making Māori research data visible and traceable; governance gives findability meaning by defining how visibility translates into authority and responsibility; and engagement ensures that both are grounded in lived relationships, tikanga, and Māori aspirations.

# 4. Concluding Remarks

In this article, we provide a brief examination of the research data landscape in New Zealand, highlighting both the opportunities and challenges inherent in the management of Māori research data. Significant gaps remain around the findability and governance of Māori research data in NZ. These issues are not just technical but are deeply entwined with questions of authority and cultural relevance. We expand on the notions of MDSov to present the idea of MRDSov to address a persistent gap between how Māori research data is produced, used, and governed in research institutions, and principles of MDSov.

To operationalise MRDSov, the article proposes an integrated approach that centres on three interdependent components: data findability, data governance, and community engagement. Data findability makes Māori research data visible and accessible in ways that respect relationships and provenance, thereby enabling informed governance. Data governance, grounded in Māori values, defines how authority and responsibility are exercised, moving away from default ownership models towards collective decision-making and accountability. Community engagement ensures that both findability and governance are not abstract concepts but are rooted in ongoing relationships, reciprocal practices, and responsiveness to Māori aspirations. Together, these pillars create a research data infrastructure that is ethical, transparent, and aligned with MDSov principles, translating institutional data commitments into practical, everyday research behaviours and policies.

Further work is needed to embed Māori Research Data Sovereignty (MRDSov) within the everyday operational structures of universities across Aotearoa New Zealand. This extends beyond symbolic policy commitments to require systemic change in how Māori research data are identified, classified, stored, accessed, and governed throughout the research data lifecycle. Achieving this will require digital research infrastructure capable of supporting culturally meaningful metadata, Māori-led governance processes, and sustained institutional practices that recognise Māori authority over Māori research data. Equally critical is the development of long-term, relational engagement strategies that enable Māori communities, iwi, and hapū to participate as decision-makers and research partners rather than as subjects of research. Together, these institutional, technical, and relational capabilities provide the foundation for operationalising MRDSov in practice, ensuring that research activities align with tikanga Māori and contribute meaningfully to Māori aspirations, knowledge systems, and collective wellbeing.

### *Acknowledgements*

This study was supported by Ngā Pae o te Māramatanga (25-28R P03 - Auaha).

# Appendix A: MDSov Principles

| MDSov Principle | Meaning in MDSov Context |
| --- | --- |
| **Rangatiratanga** (Authority) | Affirms the inherent right of Māori to exercise authority over data relating to Māori people, knowledge systems, lands, waters, and resources. Decision-making regarding the collection, access, use, interpretation, and dissemination of Māori data should rest with Māori. |
| **Whakapapa** (Connections) | Recognises that data have genealogical relationships linking them to people, communities, places, and knowledge systems. Understanding these relationships provides essential context for interpreting data and determining their appropriate use. |
| **Whanaungatanga** (Obligations) | Emphasises the relationships and reciprocal responsibilities between data holders and the Māori communities to whom data relate. Data governance should reflect ethical obligations, accountability, and ongoing relationships of care. |
| **Kotahitanga** (Collective Benefits) | Ensures that data are used to advance collective Māori aspirations and contribute to the social, cultural, environmental, and economic wellbeing of whānau, hapū, iwi, and Māori more broadly. |
| **Manaakitanga** (Reciprocity) | Requires respectful, ethical, and reciprocal data practices that uphold and protect the mana of individuals, communities, and environments throughout the data lifecycle. |
| **Kaitiakitanga** (Stewardship) | Frames Māori data as something to be cared for rather than simply owned, emphasising collective stewardship, long-term protection, and intergenerational responsibility. |

# Appendix B: Mapping MRDSov operational components with MDSov principles

| MRDSov Component | MDSov Principles | How the Principle is Upheld in Practice |
|---|---|---|
| Data Findability | Whakapapa | Metadata records provenance, origins, relationships, and context of Māori research data, ensuring data are connected to people, places, and knowledge systems. |
| | Rangatiratanga | Findability mechanisms signal Māori authority, governance expectations, and conditions of use, enabling Māori control over discovery and access. |
| | Kaitiakitanga | Visibility supports stewardship by enabling monitoring, accountability, and appropriate reuse decisions across the data lifecycle. |
| Data Governance | Rangatiratanga | Governance frameworks recognise Māori authority over decisions about access, use, interpretation, and reuse of research data. |
| | Kotahitanga | Collective decision-making structures align data governance with shared Māori aspirations and collective benefit. |
| | Manaakitanga | Governance ensures data are used ethically, responsibly, and in ways that protect Māori interests and wellbeing. |
| | Kaitiakitanga | Long-term stewardship responsibilities to current and future generations are embedded in governance arrangements. |
| Community Engagement | Whanaungatanga | Ongoing relationships between researchers, institutions, and Māori communities underpin trust and legitimacy. |
| | Kotahitanga | Engagement supports alignment between research activities, data practices, and collective Māori goals. |
| | Manaakitanga | Respectful engagement ensures Māori voices are heard, valued, and protected throughout the research process. |
| | Rangatiratanga | Engagement creates pathways for Māori communities to actively shape decisions about their research data. |